\documentclass{elsart}

\usepackage{amssymb}
\usepackage[]{graphicx}
\usepackage{multirow} \usepackage{multicol} \usepackage{colortbl} 

\newcommand{\dll}{\emph{LAM-DLL}}
\newcommand{\zemax}{\emph{ZEMAX}\circledR}
\newcommand{\lde}{\emph{Lens Data Editor}}
\newcommand{\ede}{\emph{Extra Data Editor}}
\newcommand{\mce}{\emph{Multi Configuration Editor}}

\begin{document}

\begin{frontmatter}

\title{Modeling a Slicer Mirror Using Zemax User-Defined Surface}

\author[LAM]{S. Viv\`es},
\ead{Sebastien.Vives@oamp.fr}
\author[LAM]{E. Prieto}, 
\author[LAM]{G. Moretto}, 
\author[LAM]{M. Saisse}
\address[LAM]{Laboratoire d'Astrophysique de Marseille, Site des Olives BP8-13376, Marseille, France}

\begin{abstract}
A slicer mirror is a complex surface composed by many tilted and decentered mirrors sub-surfaces. The major difficulty to model such a complex surface is the large number of parameters used to define it. The Zemax's multi-configuration mode is usually used to specify each parameters (tilts, curvatures, decenters) for each mirror sub-surface which are then considered independently. Otherwise making use of the User-Defined Surface (UDS-DLL) Zemax capability, we are able to consider the set of sub-surfaces as a whole surface. In this paper, we present such a UDS-DLL tool comparing its performance with those of the classical multi-configuration mode. In particular, we explore the use of UDS-DLL to investigate the cross-talk  due to the diffraction on the slicer array mirrors which has been a burden task when using multi-configuration mode.
\end{abstract}

\end{frontmatter}

\section{Introduction} \label{sec:intro}
Since several years a research and development activity on image slicer system for integral field spectroscopy is conducted with already in-use instrumentations, such as GEMINI/GNIRS\cite{Dubbeldam2000}, and future applications for major ground-based (VLT second-generation instruments\cite{Henault2003}) and space (JWST\cite{Prieto2003}, SNAP\cite{Ealet2002}) observatories. 

An image slicer system is usually composed of a slicer mirror array associated with rows of pupil mirrors and slit mirrors. These components are formed by a segmented assembly of several tilted and spherical mirrors. Making use of optical design software \zemax, the classical modeling method consists in using the multi-configuration mode. However, the use of such a mode implies that each mirror is independently computed compared with each other. Futhermore, such classical modeling is time-consuming because \zemax\ has to compute a large number of parameters (i.e. curvatures, tilts, decenters for each sub-mirror) and configurations (one by sub-mirror) to consider the whole instrument.

Taking advantage of UDS-DLL Zemax capability, we present an easier method to simulate segmented surfaces (slicer mirror array and rows of mirrors).

\section{Slicer Mirror User-Defined Surface} \label{sec:UDS}
For those cases where a specialized surface is required, \zemax\ supports a User-Defined Surface (UDS). All the properties of such a surface are defined in a separate C or C++ program, compiled and linked into \zemax\ using a Windows\circledR\ DLL (\textit{Dynamic Link Library}). The DLL contains functions which compute and return to \zemax\ all the data required to draw the surface, trace rays, compute refraction angles, etc.

\begin{figure}
   \begin{center}
   \begin{tabular}{c}
   \includegraphics[width=10cm]{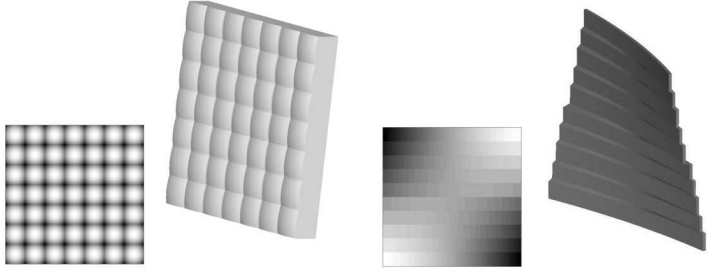}
   \end{tabular}
   \end{center}
   \caption[example] 
   { \label{fig:example} 
At left, a lens array User-Defined Surface included with \zemax\ (3D-view and surface sag). At right an example of slicer mirror array modeled using the \dll\ here developed.}
   \end{figure}

The UDS-DLL lend itself to model slicer mirrors by offering a complete description of their segmented surfaces. Fig.~\ref{fig:example} shows two complex surfaces modelled using two different UDS-DLL: a micro-lenses array as included in \zemax\ and a slicer mirror array here developed (called \dll).

\subsection{A Slicer Mirror} 
\begin{figure}
   \begin{center}
   \begin{tabular}{c}
   \includegraphics[width=10cm]{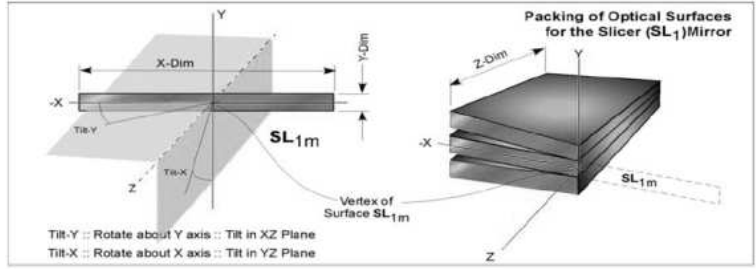} 
   \end{tabular}
   \end{center}
   \caption[notation] 
   { \label{fig:notation} 
Slicer mirror array optical specifications and notations.}
   \end{figure}

The \dll\ allows to model a slicer mirror array where each individual mirror has a rectangular clear aperture and could be spherical or flat in shape. One may specified the number of slices and their dimensions as well as curvatures and tilts for each slices. Their X- and Y-positions (Fig.~\ref{fig:notation}) are directly controlled by the \dll\ while their Z-position along the current optical axis may be specified for each slice individually. Surface sag and rays propagation are computed, and basically a close loop with \zemax\ is established in order to determine which segment of the slicer mirror array is struck by the rays on the one hand, and on the other, to use the local curvature and tilts to compute the properties of the reflected rays. 

In such a way, a slicer mirror array is correctly described by two sets of data parameters specified in the \lde\ or in the \ede\ as following described.

\subsection{Parameter data} 
Making use of \lde, the user may specified parameters defining the whole component such as the number of slices and their dimensions in both X- and Y-directions (Fig.~\ref{fig:notation}). Furthermore, three additional parameters are introduced as following:
\begin{enumerate}
	\item \textbf{"Active slice?"} defines which slice is considered in the paraxial calculation providing pupil location, magnification, effective focal length, etc.
	\item \textbf{"Centered?"} acts as a decenter of the complete surface: the optical axis may be centered on the first slice or on the entire surface of the slicer array. This parameter allows to avoid error due to the possible lack of chief ray data in the case of the number of slice is even.
	\item \textbf{"Iter."} defines the number of iterations used to compute which slice is struck by each ray. This parameter could be useful if the step between two consecutive slices is large and/or if rays come from high angles of incidence.
\end{enumerate}

\subsection{Extra data} 
Making use of \ede, the user may specified parameters defining each slice individually such as curvature, tilts in both directions (Tilt-X and Tilt-Y) and decenter along the optical axis (Z-Dec). Note that the parameter Z-Dec does not affect the whole surface but only shifts each slice's vertex along the current optical axis.

Bear in mind that the curvature is defined in the \ede\ and that only flat and spherical surfaces are considered. Thus allocations initially used by \zemax\ to specify the curvature and the conic constant in the \lde\ are ineffective. 

\section{Discussion} \label{sect:discussion}
The \dll\ acts like the commonly used "standard surface", thus optimization and analysis features of \zemax\ apply as well. The \dll\ is capable to consider all operands defined in the Merit Function making optical optimizations faster by a factor 2 to 5 for the same design using \mce. Meanwhile the construction of the Merit Function needs some adaptations to consider all facets of the segmented surface and since the \dll\ can form multiple focal spots.

  \begin{figure}
   \begin{center}
   \begin{tabular}{c}
   \includegraphics[width=10cm]{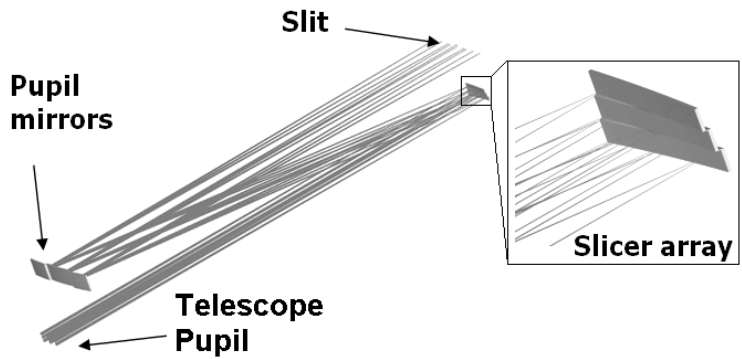} 
   \end{tabular}
   \end{center}
   \caption[design] 
   { \label{fig:design} 
Basic optical design of an IFU (3 slices). The slicer mirror array and the row of pupil mirrors are modeled by the \dll.}
   \end{figure} 

  \begin{figure}
   \begin{center}
   \begin{tabular}{c}
   \includegraphics[width=8cm]{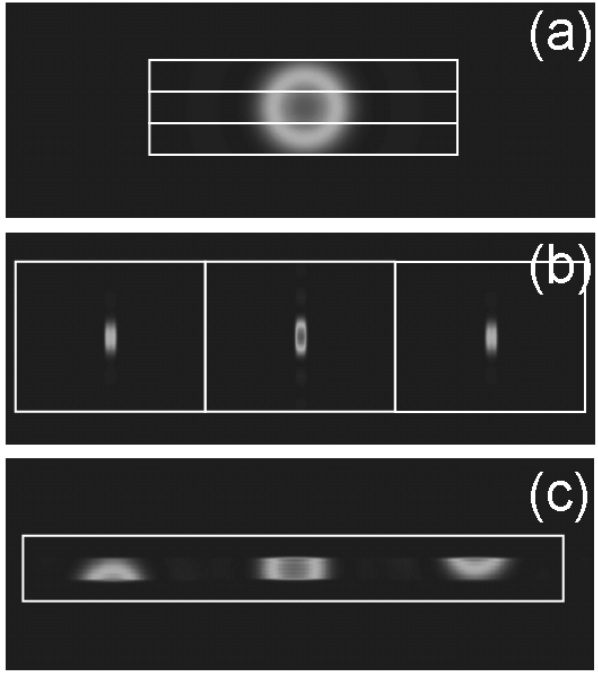} 
   \end{tabular}
   \end{center}
   \caption[cross] 
   { \label{fig:cross} 
Cross-talk analysis (diffraction). The two-dimensional point spread function (PSF) formed onto the slicer mirror array \textbf{(a)} is optically divided into three small images that are re-imaged along a slit \textbf{(c)}. The slicer mirror array has some power in order to re-image the telescope stop on the row of pupil mirrors \textbf{(b)}.}
   \end{figure} 

Analysis and tolerancing can be performed with equal results but some analysis are facilitated in particular those need to consider all slices such as the cross-talk due to diffraction. An optical layout composed by a simulated telescope, a slicer and a pupil mirror arrays (modelled by \dll) was designed in order to emphasize cross-talk due the diffraction that it is a direct issue using \dll. We used the Physical Optical Propagation (POP) tool of \zemax. The simulated telescope forms an image of a point source located to infinity onto the slicer mirrors array (Fig.~\ref{fig:cross}a). This image is optically divided into three small images that are re-formed by the pupil mirrors along a slit (Fig.~\ref{fig:cross}c). Fig.~\ref{fig:cross}b also shows the image of the telescope pupil formed by the slicer onto the row of pupil mirrors. Note that both results (Fig.~\ref{fig:cross}b and \ref{fig:cross}c) suggest some cross-talk analysis of the system.

As detailled in the section~\ref{sec:UDS}, the \dll\ uses four parameters to describe each slice that leads to 60~slices maximum since \lde\ only has 242~allocations available. This is clearly a limitation of the number of parameters describing each slice. 

Finally by making the \mce\ available, the \dll\ is appropriated to design new wide-field spectrographs using multiple integral-field units (e.g. GEMINI/GIRMOS\cite{Wright2000} which uses 32 image-slicing IFUs or VLT/MUSE\cite{Henault2003} which uses 24 IFUs).

In the immediate future, we are studying the possibilities of implementing aspherical surface shapes and variable pitch between two consecutive vertex (at the expense of reducing the maximum number of slice). 



\end{document}